# Interoperability in Planetary Research for Geospatial Data Analysis

Trent M. Hare[a], Angelo P. Rossi[b], Alessandro Frigeri[c], Chiara Marmo[d]

[a] U.S. Geological Survey, Astrogeology Science Center, 2255 N. Gemini Drive, Flagstaff, Arizona 86001
[b] Jacobs University Bremen, Campus Ring 1, 28759 Bremen, Germany.
[c] National Institute for Astrophysics, Institute for Space Astrophysics and Planetology, Viale del Parco Mellini 84, 00136 Rome, Italy.
[d] GEOPS, Univ. Paris-Sud, CNRS, Univ. Paris-Saclay, Rue du Belvédère, Bât. 509, 91405 Orsay, France

**Abstract:** For more than a decade there has been a push in the planetary science community to support interoperable methods for accessing and working with geospatial data. Common geospatial data products for planetary research include image mosaics, digital elevation or terrain models, geologic maps, geographic location databases (e.g., craters, volcanoes) or any data that can be tied to the surface of a planetary body (including moons, comets or asteroids). Several U.S. and international cartographic research institutions have converged on mapping standards that embrace standardized geospatial image formats, geologic mapping conventions, U.S. Federal Geographic Data Committee (FGDC) cartographic and metadata standards, and notably on-line mapping services as defined by the Open Geospatial Consortium (OGC). The latter includes defined standards such as the OGC Web Mapping Services (simple image maps), Web Map Tile Services (cached image tiles), Web Feature Services (feature streaming), Web Coverage Services (rich scientific data streaming), and Catalog Services for the Web (data searching and discoverability). While these standards were developed for application to Earth-based data, they can be just as valuable for planetary domain. Another initiative, called VESPA (Virtual European Solar and Planetary Access), will marry several of the above geoscience standards and astronomy-based standards as defined by International Virtual Observatory Alliance (IVOA). This work outlines the current state of interoperability initiatives in use or in the process of being researched within the planetary geospatial community.

1. Introduction

The motivation to support common, interoperable data formats, tools and delivery standards is not only to improve access for higher-level products but also to address the progressively distributed nature of ever-increasing data volumes. Terabytes of data are available for nearly all extraterrestrial bodies and some of their moons. The use of standardized formats and delivery methods have been successfully adopted by many planetary facilities. For example, the strength of using an Open Geospatial Consortium (OGC) standardized approach is that it provides consistent access to data and services that are distributed across these facilities. Data-streaming standards are well-supported by sophisticated tools used in both Geographic Information Systems (GIS) and specialized applications for the analysis of remotely sensed data. In addition, data streaming standards enable mapping applications in web browsers which not only facilitate on-line science applications but also public use (e.g., the recent EU-FP7 funded iMars web-GIS dedicated to change detection on Mars, Walter et al., 2017). Adoption of a few of these OGC standards within the Virtual European Solar and Planetary Access (VESPA) effort will not only enhance their reach but also improve the overall Virtual Observatory (VO) efficiency by upgrading existing VO standards to adapt them for planetary sciences. A major goal for VESPA will be the addition of these geoscience capabilities in existing VO tools providing alternative but popular data mining, retrieval, advanced visualization and data analysis already available in many of the more astronomy-based VO tools (Erard et al., 2014). Herein, we provide an update of planetary geospatial interoperability initiatives, and examples of their successful application.

The most obvious difference between extraterrestrial data sets and terrestrial data sets is simply the shape or size of the planetary body. For example, nearly all larger bodies in our solar system have defined geodetic parameters, documented by the International Astronomical Union (IAU), allowing capable mapping applications to study these bodies. Recognizing the need for standardized geodetic control on planetary bodies, the IAU established the Working Group on the Cartographic Coordinates and Rotational Elements of Planets and Satellites in 1976 (Archinal et al., 2011). This group reports every 3 to 5 years on the preferred rotation rates, spin axes, prime meridians, and reference surfaces for planets and satellites, which helps ensure that digital mapping endeavors are effectively comparable. This planetary standards group provides the critical foundation for the implementation of all the initiatives described below.

Discussed in this report are some of the different interoperable initiatives within the planetary community including interoperable raster formats and tools, web mapping standards, the communication of coordinate reference systems, cartographic mapping standards, and data portals.

## 2. Interoperable Initiatives

### 2.1 Interoperable Raster Formats

Most planetary data acquired by both NASA and non-US spacecraft are archived in a Planetary Data System (PDS) format (McMahon, 1994). The PDS is managed by NASA Headquarters' Planetary Sciences Division and consists of a collection of external facilities to support archiving and distribution of planetary data. The bulk of the PDS data holdings are cataloged in their original raw instrument form, however, to best use these data sets within GIS applications, they should first be spatially referenced to the planetary body. Unfortunately, the PDS format, while well-documented with its focus on long-term availability of the archives, is not widely recognized by mapping applications although support is improving. Two formats which have been targeted for their planetary support and are now commonly used in the community include GeoTIFF and GeoJPEG2000.

*GeoTIFF*: Probably the most popular geospatial format is GeoTIFF. The GeoTIFF format, fully within the public domain, was created by Dr. Niles Ritter in the 1990s during his term at the Jet Propulsion Laboratory (Ritter and Ruth, 2000). GeoTIFF makes use of geospatial tags embedded within the TIFF file format. It is one of the only image formats which allow the flexibility to support tag structures without causing issues for applications that do not support those tags. The image format can support 8-bit grayscale images, and up to 16, 32 and 64-bit floating point elevation models. TIFF also supports a variety of compression and tiling options to increase the efficiency of image reading and online distribution. The BigTIFF extension allows single images to be greater than 4 gigabytes.

*GeoJPEG2000*: Although support might be retracted for version 4 of the standard, PDS has approved the use of the JPEG2000 format in the PDS3 standard. This format supports the exact same tags as the GeoTIFF format but it is stored within a Universally Unique Identifier (UUID) container. When utilized, this format is informally called GeoJPEG2000 (also GeoJP2™). In 2008, University of Arizona's Mars HiRISE instrument team was the first mission to release their map-projected PDS archives using a hybrid method combining the use of the GeoJPEG2000 standard and a detached PDS label (McEwen et. al., 2007). The simple text PDS label is necessary to hold required PDS metadata, for example, author, instrument particulars, or mission dates which are not suitable for the geospatial container.

While this hybrid approach (detached PDS label and imbedded geospatial container) sounds like the best of both worlds, the JPEG2000 format does not have broad support for 32-bit floating point values. It is also unfortunate that the open Jpeg2000 libraries (e.g., OpenJPEG or Jasper) are still lacking in capabilities and speed to proprietary solutions like the Kakadu library (http://kakadusoftware.com/).

*GeoFITS:* The Flexible Image Transport System (FITS) has been defined for data acquisition and archiving in astronomical observatories and is used for spatial telescope data. FITS is one of the standard formats in the Virtual Observatory (VO). It is compatible with PDS archiving specifications and is supported by a large number of open libraries and software tools. The format has benefited from the well-established CFITSIO library (Pence, 1999) which provides simple high-level routines for reading and writing FITS files and is today well supported by the community driven software development project AstroPy (http://www.astropy.org/). Unfortunately, this format is not currently widely used within the planetary domain but there are initiatives to extend the FITS standard to support geospatial tags. This update to the standard, informally called GeoFITS (Marmo et al., 2016), could be an opportunity to allow more seamless sharing of data across the astronomy and planetary domains and potentially homogenize methods from acquisition to visualization.

### 2.2 Interoperable GDAL-based Tools

The U.S. Geological Survey's Astrogeology Science Center (ASC) is a major contributor of software for cartographic data processing in support of National Aeronautics and Space Administration (NASA) missions and research programs, including the Planetary Program, Code S flight projects, research and data analysis projects, and the PDS (Hare et al., 2014). ASC supports the Integrated Software for Imagers and Spectrometers (version 3, ISIS3), a specialized image processing package for working with planetary image data (Keszthelyi et al., 2014). While ISIS3 can ingest and export several different formats, it is only able to process in its own specialized image format. In 2006 and 2007, PDS and ISIS3 format support was added into the Geospatial Data Abstraction Library (GDAL) to improve interoperability with other applications. This added capability to GDAL opened up these very planetary-specific formats for conversion and more importantly direct use within several mapping applications.

*GDAL*: GDAL, released by the Open Source Geospatial Foundation (OSGeo), offers powerful capabilities for converting and processing planetary data. GDAL is a format translation library for geospatial raster and vector data (GDAL, 2016). In addition to the aforementioned PDS and ISIS3 reader, GDAL also supports other planetary formats including ISIS2, and more recently the Video Image Communication and Retrieval (VICAR) format (Walter and van Gasselt, 2014). Mapping applications, which utilize the GDAL library for raster I/O (Input/Output), can directly access these formats. This has greatly reduced the need to standardize on a single format. And with GDAL's support for dozens of additional formats, it has allowed the community to more easily collaborate across groups that may prefer to work in differing formats, either due to their preference or software requirements. For applications that do not use GDAL for I/O, the bundled routines released with GDAL can be used to convert these formats into more universal geospatial formats (e.g., GeoTIFF).

*Desktop GIS and GDAL*: Some popular geospatial applications with GDAL support include QGIS, GRASS GIS, gvSIG, Esri's ArcMap and ArcGIS Pro, and Generic Mapping Tools (GMT). While Esri's applications are commercial, all the others are Free Open Source (FOSS) software. In particular, GRASS GIS and GMT implement a very large number of tools for geoprocessing. Similarly, gvSIG and QGIS contain a wide range of geoprocessing tools but also feature user friendly graphical user interfaces. ArcMap and ArcGIS Pro support both geoprocessing and a graphical user interface and additionally support the required symbologies for proper geologic map publication (see section 2.5). It is not that the other applications cannot support these symbologies but no one has invested the time to convert the needed symbologies for geologic maps. Fortunately, an initiative is underway to import these symbologies for QGIS but the conversion can be tedious (http://github.com/afrigeri/geologic-symbols). Alongside these excellent desktop GISs, it is worth mentioning Arizona State University's JMARS as a streaming desktop GIS built specifically for the planetary domain (Christensen et al., 2009). While JMARS indirectly uses GDAL for some format support, most data layers are streamed using their own tile-cache protocols but their back-end server also supports the OGC WMS protocol. JMARS has established itself as an important tool for many researchers and as a great introductory GIS application for planetary educators.

*Scripting Languages and GDAL*: While GDAL is written in C/C++, it has bindings for use with many languages, including JAVA, PERL, Python, and .NET. As an interoperability example, the authors highlight Python, which has a robust standard library and mature scientific computing stack (e.g., Numerical Python (NumPy), Scientific Python (SciPy), Pandas, Matplotlib). GDAL provides the interface to support data reads into a common, in-memory format, known as the NumPy n-dimensional array. This opens a world of extremely powerful image processing and analytical methods. ASC utilizes Python for both rapid prototyping and production development. For example, to support the NASA's InSight Mars lander and Mars 2020 rover missions, specialized topographic slope software was being supported in an outdated code-base. Using GDAL, Python, and existing array filtering functions in SciPy, the original source code was quickly ported and integrated within the digital terrain model workflow. During the port, histogram binning was added (using NumPy) to combine histogram and cumulative slope graphs (using Matplotlib) and create colorized slope figures to assist in the ability to land the spacecraft safely on the surface (Figure 1).

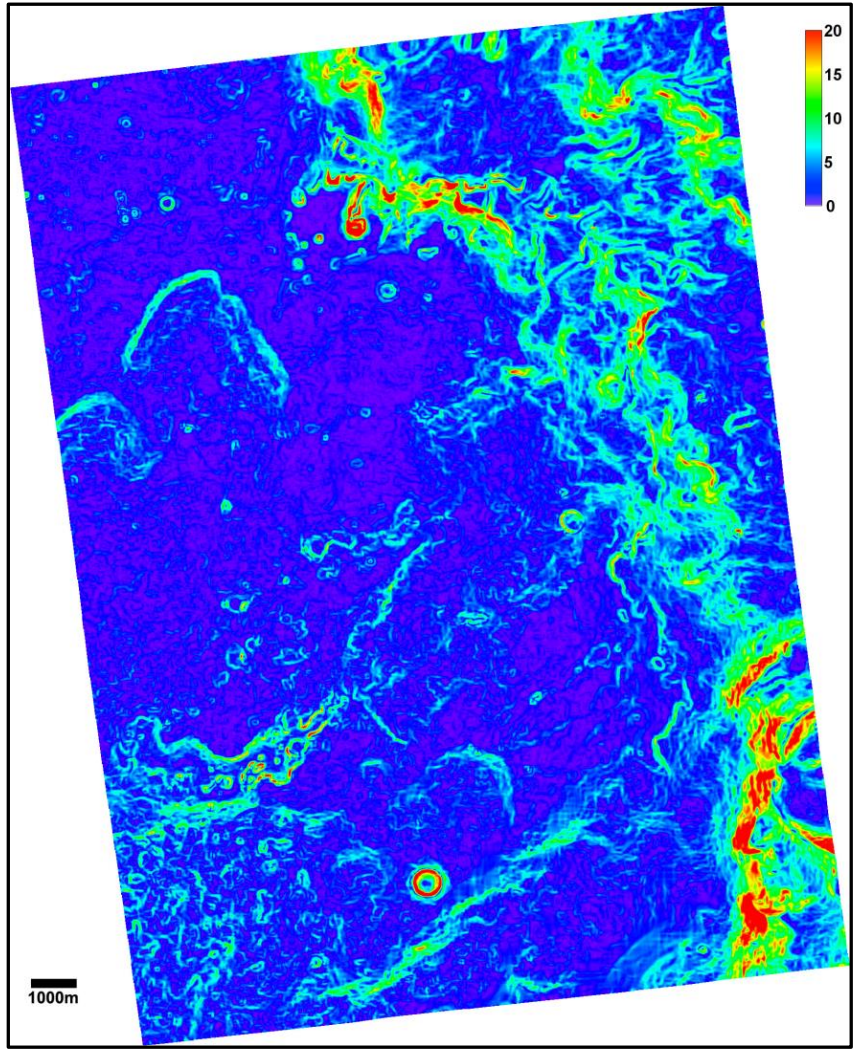

Figure 1: Example derived slope map at 20 meters per pixel within McLaughlin crater (center at 21.9° north 337.63° east) generated for the Mars 2020 rover mission. The slope map will be used to help assess the ability to land the rover safely on the surface. Image credit: NASA/USGS.

*2.3 Interoperable Web Services*

The OGC is a consortium of more than 500 international companies, universities, and government agencies which define standards such as the Web Mapping Services (simple image maps), Web Map Tile Services (WMTS, cached image tiles), Web Feature Services (WFS, feature streaming), Web Coverage Services (WCS, rich scientific data streaming), and Catalog Services for the Web (CSW, data searching and discoverability).

*WMS/WFS Services*: In short, a WMS accepts queries for map-projected layers and returns requested data generally in the form of a simple image and well-known graphical format (e.g., JPEG, PNG). A WFS returns geographical features including spatial vector geometries (point, line, or polygon) and their attributes such as feature name, feature type, or any associated measures (e.g., length, area). ASC supports both WMS and WFS allowing capable mapping clients to view full-resolution base maps and geospatial databases. ASC currently supports more than 100 global and polar image layers across 30 different planetary bodies and moons (http://bit.ly/AstroWMS). For ArcMap GIS users, these layers are listed on Esri's ArcGIS Online data portal under the Planetary GIS group (http://bit.ly/PlanetaryGIS). Several other facilities maintain planetary WMS servers and some include support for the proposed IAU planetary projection codes as described below (e.g., Lunaserv by Arizona State University, Estes et al., 2013).

*WMTS Services:* WMTS quickly gained popularity as the need to more quickly deliver image layers grew. It is well-known that WMS can be somewhat compute and disk intensive. WMTS introduced a method for the client to request a defined pattern of pre-generated tiled and generally cached images. These image caches can not only be locally held on the server but can also be cached on the client for immediate access to areas already visited. By first preparing these tiles prior to access, it allows a relatively modest server to support extremely fast layer requests. Although WMTS is fast, it is not as flexible as WMS which can generate images at any scale. This means the client application needs to either lock into the WMTS pre-rendered scales or resample the tiles to the scale requested by the user. As an example, the Jet Propulsion Laboratory and NASA Ames Research Center are now actively supporting numerous WMTS services for their web-based mapping applications called MarsTrek (Figure 2), VestaTrek and MoonTrek (Day and Law, 2016).

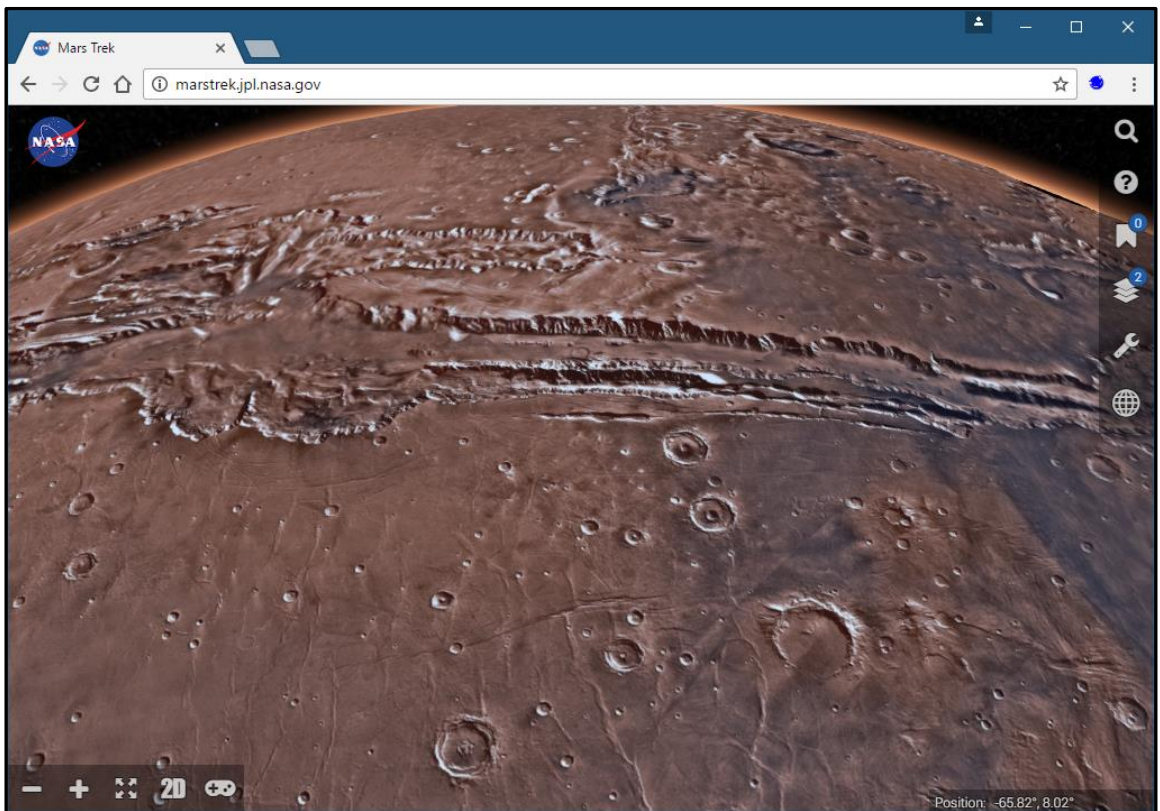

Figure 2: This browser window shows the MarsTrek on-line mapping interface which is backed by an OGC WMTS tile-cache image server. Shown is the Mars Viking-based colorized digital image mosaic version 2.1 draped over the Mars Orbiter Laser Altimeter elevation model (NASA/JPL/USGS). http://marstrek.jpl.nasa.gov.

*WCS/WCPS Services:* Distinct from WMS services, which generally return simple image maps, the WCS provides a method to stream "data" (also called raster coverages). For example, a WMS will generally return a grayscale representation of a digital elevation model but a WCS will actually return a raster coverage with elevation values intact. Here a simple JPEG-formatted image cannot be used and will require a client to use a more capable format like GeoTIFF. It should be noted that a WMS can also return raster coverage data formats like GeoTIFF, however most clients may have trouble displaying it. Alongside WCS, Web Coverage Processing Services (WCPS) are generally used. The WCPS standard defines a language for filter and/or processing these raster coverages prior to shipping to the client.

 A community example of an OGC WCPS is PlanetServer, developed at Jacobs University Bremen, Germany (Figuera et al., 2015). PlanetServer is comprised of two sides: the server side database called Rasdaman (Baumann et al. 2016) and the client side viewer based on NASA's Web World Wind. Web World Wind is a general-purpose 3D/4D client used as a virtual globe to interactively analyze and visualize hyper-spectral and elevation data. PlanetServer has also implemented the WCPS protocol to support communication between client and server.

*2.4 Coordinate Reference Systems*

Hare et al. (2006) proposed methods to support planetary Coordinate Reference Systems (CRS) within existing OGC web mapping standards. Within a typical OGC web mapping session, the server must define a minimum set of information such that the client application understands not only the data layer but also the current CRS and/or map projection. Generally, web mapping servers default to using the numeric European Petroleum Survey Group (EPSG) codes to define the CRS or Spatial Reference System (SRS).  For example, code "4326" is the EPSG identifier for Earth's "WGS 84" geographic CRS.  The server and client relay this code by passing an in-line SRS request using the string "SRS=nameSpace:code" (e.g., "SRS=EPSG:4326"). Additional EPSG codes were generated to attempt to catalog the most widely used cartographic map series from all countries (e.g., "32612" = WGS 84 / UTM zone 12N; "21413" = Beijing 1954 / Gauss-Kruger zone 13).  However, if a CRS is not part of the EPSG database, and no planetary definitions are present, there is not an easy mechanism to explicitly define custom settings within the EPSG namespace

To help solve this incompatibility among planetary servers, the authors have proposed their own set of codes outside of the EPSG namespace but for eventual adoption by the OGC. The proposed codes will leverage both the IAU's publication date for their Report on Cartographic Coordinates and Rotational Elements (Archinal, et al., 2011) and the Navigation and Information Facility (NAIF, http://naif.jpl.nasa.gov) numeric code for extraterrestrial bodies.

First, given the three- to five-year publication cadence used by the IAU, the required namespace is defined using an alpha-numeric code (e.g., IAU1979, IAU2000, IAU2009). Extracting the year from the namespace may be necessary to allow the year to be used as a separate version number. The alpha-numeric code specification allows for updates to the definition of the body, which is common as better data are gathered, to coincide with updates to the IAU publication.

Second, the numeric codes will be modeled after the existing NAIF system. The NAIF system defines the barycenter (center of mass) of the solar system as 0 and defines the Sun as 10. This allows the planets to be classified as 1 through 9 starting with Mercury out to Pluto. The NAIF planet ID is then defined as the planet barycenter ID * 100 + 99. Thus Mars, in the NAIF system, is defined as "499". To build upon that value, the new geospatial code for Mars is derived as follows:

- Planetary IAU2000 code = 499 * 100 = 49900
- Sample WMS call: SRS="IAU2000:49900"

The moons for each body, as defined by NAIF, start at planet barycenter ID * 100 + 1. For example, Phobos is defined as "401" and Deimos as "402". The new planetary code would be defined as Deimos IAU2000 code = 402* 100 = 40200.

To continue with the Mars example, the first 10 numbers, 49900 to 49909, are reserved for geoid definitions (Table 1). Codes in the range of 49910 to 49959 are reserved for predefined projection definitions intended to capture the most popular projections used in the planetary community. Codes from 49960 to 49999 are for "AUTO" projections. AUTO projections allow the user to also submit the projection parameters (e.g., SRS="IAU2000:49964,9001,100,45", where 49964 is Transverse Mercator, 9001 is the EPSG code for meters, center longitude=100º and center latitude=45º).

Table 1: Example planetary codes to support planetary WMS servers for Mars using the "IAU2000" namespace. Other bodies will follow similar definitions as derived from the NAIF planetary codes. Tables for the codes and Python scripts to generate the codes are available at Github (http://bit.ly/IAU2000).

| IAU Name | Mars IAU | GEOIDS |
|---|---|---|
| IAU2000 | 49900 | Mars2000, aerocentric latitudes, positive East longitudes |
| IAU2000 | 49901 | Mars2000, aerographic latitudes, positive West longitudes |
| IAU2000 | 49902 - 49909 | Available |
| | | **PROJECTIONS - Even codes=aerocentric, Odd codes=aerographic** |
| IAU2000 | 49910 | **Equirectangular (Simple Cyl), clon=0º, spherical equation, areocentric** |
| IAU2000 | 49911 | **Equirectangular (Simple Cyl), clon=0º, spherical equation, aerographic** |
| IAU2000 | 49912 | **Equirectangular (Simple Cyl), clon=180º, spherical equation, areocentric** |
| IAU2000 | 49914 | **Sinusoidal, clon = 0º, spherical equation, areocentric** |
| IAU2000 | 49916 | **Sinusoidal, clon = 180º, spherical equation, areocentric** |
| IAU2000 | 49918 | **Polar Stereographic, clat=90º, clon = 0º, spherical equation, polar radius** |
| IAU2000 | 49920 | **Polar Stereographic, clat=-90º, clon = 0º, spherical equation, polar radius** |
| IAU2000 | 49922 ~ 49959 | Available (1:2M Mars series handled by AUTO below) |
| | | **AUTO PROJECTIONS (parameter order)** |
| IAU2000 or Auto | 49960 | Auto Sinusoidal, spherical equation, areocentric, (clon) |
| IAU2000 or Auto | 49961 | Auto Sinusoidal, spherical equation, aerographic, (clon) |
| IAU2000 or Auto | 49962 | Auto (Polar) Stereographic, spherical equation, (clon, clat, scale) |
| IAU2000 or Auto | 49964 | Auto Transverse Mercator, areocentric, (clon, clat, scale) |
| IAU2000 or Auto | 49966 | Auto Orthographic, spherical equation, areocentric, (clon, clat) |

As mentioned above, the EPSG coded system has a narrow set of predefined CRSs. To help address this, the OGC has begun efforts to support and extend EPSG codes with a parametric URL-based CRS scheme called SECORE (Semantic Coordinate Reference System Resolver, Rossi et al., 2016). Such a system catalogs and accepts an HTTP parametrized CRS as input and returns a CRS definition formatted using a verbose Geography Markup Language (GML) definition. The authors plan to extend and implement the IAU code-set within SECORE. Once available, these definitions can be dynamically converted to other formats, such as Proj4 ([https://trac.osgeo.org/proj/](https://trac.osgeo.org/proj/)) or the "well-known text" (WKT) map projection OGC standard.

While these codes have existed for several years, it is still recommended that a planetary web service also maintains support for the decimal degree (latitude/longitude) Earth-based code "EPSG:4326", to ensure compatibility across software and online viewers. The use of this Earth-base code will allow nearly all (Earth-based and planetary) clients to render data from these WMS services but issues in measurement can arise due to assumptions made from using an incorrect spheroid definition. It is worth noting, that more capable applications like QGIS and ArcMap which can reproject these WMS image services on-the-fly to a properly defined planetary map projection, will function correctly using this work-around.

*2.5 Cartographic Interoperability*

As described above, the IAU defines the recommended rotation rate, spin axis, prime meridian, and reference surface for individual planets and satellites; however, their oversight does not cover other standards essential for digital mapping such as common feature attributions, feature symbols, recommended mapping scales, and metadata. When possible, it is suggested that digital maps use these standards so that consistent map products can be developed.

Feature attributes and their assigned symbols for planetary digital maps are commonly defined in the Digital Cartographic Standard for Geologic Map Symbolization (Skinner et al., 2016) prepared by the USGS for the Federal Geographic Data Committee (FGDC). For example, recommended attributes for geologic contacts or geologic unit boundaries include attributes for contact certainty. The nominal level categories are then given explicit symbolic representations such as solid black lines for certain contacts or dashed black lines for approximate boundaries. Cartographic symbologies are primarily drawn from the same sets of attributes and symbols as used for Earth. This heritage facilitates an understanding and readability of geologic or thematic planetary maps because readers are familiar with the feature attribution names and familiar symbol styles (Nass et al., 2010).

*2.6 Interoperable Metadata and Data Portals*

*Metadata*: In short, metadata is the ancillary documentation that helps describe the rationale, authorship, attribute descriptions, spatial reference, and other pertinent information for data. For planetary data, PDS archives are the required method to document data products. This is in contrast to most United States Earth-based geospatial archives which mandate use of the FGDC or the related International Organization for Standardization (ISO) metadata standard. Methods for conversion from a PDS standard to a FGDC/ISO standard should be possible, given that the FGDC/ISO standards only require a few minor additions to properly support planetary data (Hare, 2011). This is an important translation to support so that existing FGDC/ISO tools can be used for planetary data.

Metadata is not only a United States Federal mandate; it is also a priority focus for the European Commission, the European Union's executive body.

*Data Portals:* One of the latest trends in the geospatial community, including the planetary community, is to provide data portals (e.g., http://www.data.gov/). These portals assemble data collections for online browsing and download. Many Earth-based data portals are built around the use of FGDC/ISO metadata to import, describe, and catalog data for external users. Most existing planetary data portals provide easy access to a data collection for browsing and retrieval but they often include minimal metadata and thus have limited search capabilities for outside users. Methods defined by the OGC CSW standard will facilitate such outside access, so that users need not build new search tools or application layer interfaces (Figure 3).

One major benefit of using the OGC CSW standard is that portals can support searches across data catalogs, because the standard allows one data portal to index data in another. Products served by such mutually indexed portals will need to give appropriate credit for and references to the data creators including pointing back to the original host data portal. In summary, the benefits for implementing a data portal using the OGC CSW standards include

- enabling easy search and discovery of existing geospatial data and services;
- reducing redundancy across portals; and
- establishing authoritative versions.

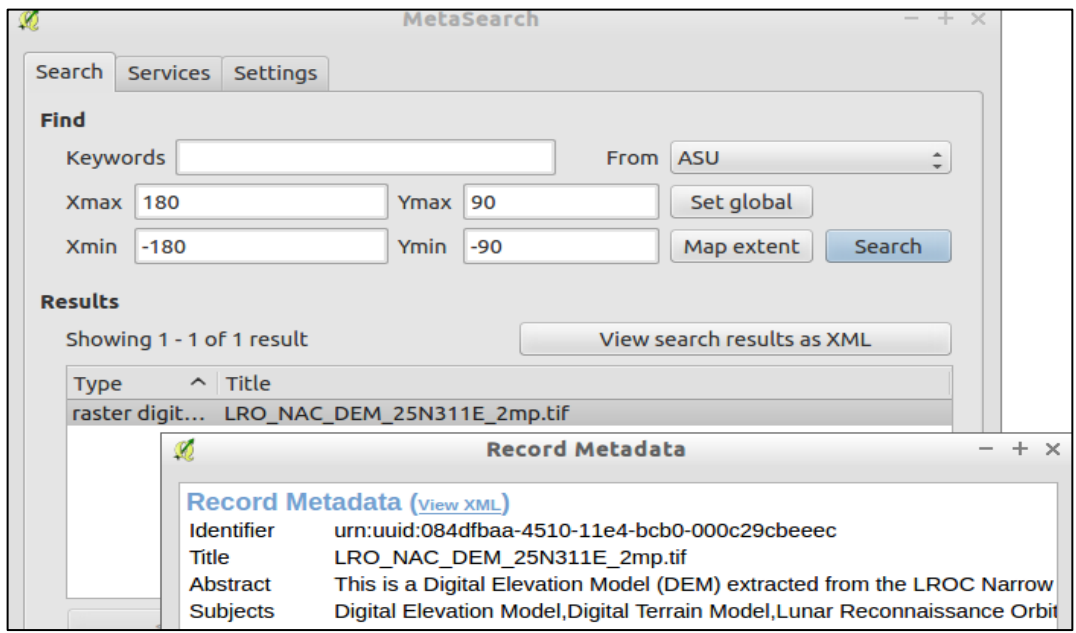

Figure 3: An example CSW search using the Desktop GIS application QGIS with a CSW plug-in. In this case, the CSW server returned a digital elevation model as derived from Lunar Reconnaissance Orbiter Camera images.

For many years, the astronomy community has been successfully sharing data via similar data portals called the Virtual Observatories (VO). The VO portals contain mature cataloging and searching methods. Leveraging these robust methods for planetary science will enhance data discoverability across the different but related astronomy field. The VESPA initiative will use these VO standards with the goal for a tight integration of astronomy data and planetary science data to provide a very efficient cross-pollination for neighboring domains (Erard et al., 2014).

## 3. Discussion

While advances have been made concerning planetary geospatial interoperable initiatives, many not discussed here, there are still plenty of issues to tackle. The adoption of OGC standards remain a challenge due to their primary creation for Earth-based applications and for some specific standards simply because they are complicated and do not yet have broad support across geospatial applications. Straightforward standards like WMS have been well suited for mass acceptation but the capabilities are limited. More capable but complicated and data-intensive standards like WCS will always take more time for adoption. And initiatives like VESPA are critical to help bring and incorporate these geoscience standards for use within the astronomy domain to better align these two disparate but related fields.

As a community, we need to limit our use of "work-arounds" within existing standards and make sure planetary data and CRSs are supported as a core feature within a standard and within applications. Fortunately, many formats, web services, and applications are fully capable of correctly using planetary data. And when they have limitations, many across our community are quickly taking on their incompatibilities. ASC and others have worked with Esri on planetary support within ArcMap and more recently ArcGIS Online. Planetary updates have been published for the widely-used online map viewer OpenLayers (https://pilot.wr.usgs.gov). The online globe-viewers Cesium and Nasa World Wind JavaScript are actively being updated to support planetary data (https://github.com/epn-vespa/cesium, https://github.com/planetserver). Due to the efforts of the Lunar Reconnaissance Orbiter Camera Team at Arizona State University, the proposed IAU2000 projection codes now ship with QGIS and efforts by co-author Dr. Frigeri, these radius definitions are also included within the GRASS environment (Frigeri et al., 2011).

Lastly, many other standards, formats and tools will require continued research, and if needed, modified to support the planetary domain. This includes OGC standards like Web Processing Services (WPS), new formats like GeoPackage, and the geospatial extensions for OpenSearch (as released by Amazon) just to name a few. More infrastructural technologies like the PostgreSQL/PostGIS relational database management system and geospatial servers like GeoServer and MapServer will continue to require updates to better support planetary services. And even if these tools can already handle planetary data or projections, resources and tutorials should be published to help invite broader adoption.

## 4. Conclusions

As data volumes grow, interoperable methods of accessing and working with geospatial data will continue to be critical to allow the numerous technical and scientific fields in the planetary community to work together. Standardized methods for direct access to on-line planetary data will also continue to rapidly mature. But standards are only worth implementing if they are actually adopted by researchers and across facilities. Thus, the authors will continue to encourage more facilities to use OGC and VO interoperable standards for distributing and hosting data sets. OGC- and VO-based technologies have proven to handle very diverse data sets, like terabyte-sized mosaics, hyper-spectral imagery, and high-resolution non-continuous images and simultaneously preserve the ease and accessibility for mapping and research applications.

**Acknowledgments**

The research described in this paper was supported by funding through the National Aeronautics and Space Administration and the European Commission's Horizon 2020 program. Sebastian Walter (Freie Universität Berlin) and one anonymous reviewer provided significant feedback for this manuscript. Finally, we thank the many Astrogeology staff at the U.S. Geological Survey and other planetary facilities who have worked on and collaborated with these initiatives. None of the topics described above can live without the support and adoption by others. Any use of trade, firm, or product names is for descriptive purposes only and does not imply endorsement by the U.S. Government.

# References

Archinal, B.A., A'Hearn, M.F. Bowell, E., Conrad, A., Consolmagno, G.J., Courtin, R., Fukushima, T., Hestroffer, D., Hilton, J.L. Krasinsky, G.A., Neumann, G., Oberst, J., Seidelmann, P.K., Stooke, P., Tholen, D. J., Thomas, P.C., Williams, I. P. (2011) Report of the IAU Working Group on Cartographic Coordinates and Rotational Elements: 2009, *Celestial Mechanics and Dynamical Astronomy*, 109, no. 2, February, 101-135, doi:10.1007/s10569-010-9320-4.

Baumann, P. Merticariu, V. Dumitru, A. Misev, D. (2016) Standards-Based Services for Big Spatio-Temporal Data, ISPRS - International Archives of the Photogrammetry, *Remote Sensing and Spatial Information Sciences*, Volume XLI-B4, 2016, pp.691-699.

Christensen, P.R. Engle, E. Anwar, S. Dickenshied, S. Noss, D. Gorelick, N. Weiss-Malik, M. (2009) JMARS – A Planetary GIS, *American Geophysical Union*, Fall Meeting abstract #IN22A-06.

Day, B. H., Law, E. S. (2016) Education and Engagement Applications of NASA Lunar and Planetary Mapping and Modeling, Lunar Planet Science Conference LXXVIII. *Lunar and Planetary Institute*, Houston, TX, abstract #1903.

Edmundson, K. L., Cook, D.A., Thomas, O.H., Archinal, B.A., Kirk, R.L. (2012) Jigsaw: the ISIS3 Bundle Adjustment for Extraterrestrial Photogrammetry, ISPRS Annals of Photogrammetry, *Remote Sensing and Spatial Information Sciences*, Volume I-4, 2012, pp.203-208.

Erard S., P. Le Sidaner, B. Cecconi, J. Berthier, F. Henry, C. Chauvin, N. André, V. Génot, C. Jacquey, M. Gangloff, N. Bourrel, B. Schmitt, T. Capria, G. Chanteur (2014) Planetary Science Virtual Observatory architecture. Astronomy & Computing 7-8, 71-80. doi: 10.1016/j.ascom.2014.07.005.

Estes, N.M., Hanger, C.D., Licht, A.A., Bowman-Cisneros, E. (2013) Lunaserv Web Map Service: History, Implementation Details, Development, and Uses, Lunar and Planetary Science Conference XLIV, *Lunar and Planetary Institute*, Houston, TX, Abstract #1719.

Frigeri A., Hare, T., Neteler, M., Coradini, A., Federic, C., Orosei, R., 2011, A working environment for digital planetary data processing and mapping using ISIS and GRASS GIS: *Planetary and Space Science*, v. 59, p. 1265–1272.

Figuera, R.M., Rossi, A.P., Baumann, P., Pham, H.B., Misev, D., Merticariu, V., Dumitru, A., Hogan, P. (2015) Analyzing lunar DTMs through web services with EarthServer/PlanetServer-2, *International Society for Photogrammetry and Remote Sensing Working Group IV/8 Meeting*, Berlin, Germany.

GDAL (2016) GDAL - Geospatial Data Abstraction Library: Version 2.0.1, *Open Source Geospatial Foundation*, http://gdal.osgeo.org Last visited 6/30/2016.

Greeley, Ron, Batson, R.M. (1990) Planetary Mapping, Cambridge Planetary Science Old, *Cambridge University Press*, ISBN 9780521033732.

Hall, R. C., Lunar Impact: A History of Project Ranger, *NASA SP-4210* (1977) URL: http://history.nasa.gov/SP-4210/pages/Cover.htm Last visited 6/30/2016.

Hare, T.M., Archinal, B., Plesea, L., Dobinson, E., Curkendall, D. (2006) Standards Proposal to Support Planetary Coordinate Reference Systems in Open Geospatial Web Services and Geospatial Applications, Lunar Planetary Science Conference XXXVII, *Lunar and Planetary Institute*, Houston, TX, Abstract #1931. Github URL: http://bit.ly/IAU2000.

Hare, T.M. (2011) Standards-Based Collation Tools for Geospatial Metadata in Support of the Planetary Domain (thesis), *Northern Arizona University*, Flagstaff, AZ.

Hare, T.M., Gaddis, L.R., Keszthelyi, L.P. (2014) Online Planetary Data and Services at USGS Astrogeology, Lunar and Planetary Science Conference XLV, *Lunar and Planetary Institute*, Houston, TX, Abstract #2487.

Heacock, R. L., Kuiper, G. P., Shoemaker, E. M., Urey, H. C., Whitaker, E. A. (1965) Ranger VII, Part II, Experimenters' Analyses and Interpretations, Technical Report No. 32-700, *Jet Propulsion Laboratory,* Pasadena, CA.

Keszthelyi, L., Becker, T., Titus, T., Sides, S., Gaddis, L., Hare, T., Kirk, R., Edmundson, K., Anderson, J. (2014) Utilizing the Integrated Software for Imagers and Spectrometers (ISIS) to Support Future Missions, Lunar Planetary Science Conference XLV, *Lunar and Planetary Institute*, Houston, TX, Abstract #1686.

Madden, Margaret, ed. (2009) Manual of Geographic Information Systems, *American Society for Photogrammetry and Remote Sensing*, ISBN 9781570830860.

Marmo, C., Hare, T. M., Erard, S., Cecconi, B., Costard, F., Schmidt, F., Rossi, A.P. (2016) FITS Format for Planetary Surfaces: Bridging the Gap Between FITS World Coordinate Systems and Geographical Information Systems, Lunar Planet Science Conference LXXVIII. *Lunar and Planetary Institute*, Houston, TX, abstract #1870.

McEwen, A. S., Eliason, E. M., Bergstrom, J. W., Bridges, N. T., Hansen, C. J., Delamere, W. A., Grant, J. A., Gulick, V. C., Herkenhoff, K. E., Keszthelyi, L., Kirk, R. L., Mellon, M. T., Squyres, S. W., Thomas, N., Weitz, C. M., (2007), Mars Reconnaissance Orbiter's High Resolution Imaging Science Experiment (HiRISE), J. Geophys. Res., 112, E05S02, doi:10.1029/2005JE002605.McMahon, S.K. (1994) Overview of the Planetary Data System, *Planetary and Space Science*, volume 44, #1, pp. 3-12.

Nass, Andrea, van Gasselt, S., Jaumann, R, Asche, S. (2010) Implementation of cartographic symbols for planetary mapping in geographic information systems. *Planetary Space Science*, doi:10.1016/j.pss.2010.08.022.

Pence, W. (1999) in ASP Conf. Ser., Vol. 172, Astronomical Data Analysis Software and Systems VIII, ed. D. Mehringer, R. Plante, and D. Roberts (San Francisco: ASP), 487.

Ritter, N., Ruth, M. (2000) GeoTIFF Format Specification, Revision 1.0, Version 1.8.2, December https://trac.osgeo.org/geotiff/.

Rossi, A.P., Hare, T., Baumann, P., Misev, D., Marmo, C., Erard, S., Cecconi, B., Marco Figuera, R. (2016) Planetary Coordinate Reference Systems for OGC Web Services, Lunar Planetary Science Conference XLVII, *Lunar and Planetary Institute*, Houston, TX, Abstract #1422.

Schimerman, L.A., ed. (1973) Lunar Cartographic Dossier, Volume I, *NASA and the Defense Mapping Agency*, St. Louis, Missouri. with changes and additions to 1977 July. (approx. 359 pp). http://www.lpi.usra.edu/lunar_resources/lc_dossier.pdf.

Skinner, J.A., Jr., Hare, T.M., Fortezzo, C.M., and Hunter, M.A. (2016) Planetary Geologic Mapping Handbook., U.S. Geological Survey. *Astrogeology Science Center,* Flagstaff, AZ, http://planetarymapping.wr.usgs.gov/Page/view/Guidelines.

Walter, S. H. G., van Gasselt, S. (2014) HRSC Data Dissemination - Dynamic Queries and Data Interoperability, Lunar Planetary Science Conference XLV, *Lunar and Planetary Institute*, Houston, TX, Abstract #1088.

Walter, S. H. G., Steikert, R., Schreiner, B., Muller, J.-P., van Gasselt, S., Sidiropoulos, P., Lanz-Kroechert, J. (2017) The iMars webGIS – Space-Time Queries and Dynamic Time Series of Single Images, Lunar Planetary Science Conference XLVIII, *Lunar and Planetary Institute*, Houston, TX, Abstract #1066.